**Linking reduced breaking crest speeds to unsteady nonlinear water wave group behavior**

M.L. Banner[1]*, X. Barthelemy[1,2], F. Fedele[3], M. Allis[2], A. Benetazzo[4], F. Dias[5] and W.L. Peirson[2].

1. School of Mathematics and Statistics, The University of New South Wales, Sydney 2052, Australia

2. Water Research Laboratory, School of Civil and Environmental Engineering, The University of New South Wales, Manly Vale, NSW 2093, Australia

3. School of Civil and Environmental Engineering, and School of Electrical and Computer Engineering, Georgia Institute of Technology, Atlanta, GA 30332, USA.

4. Institute of Marine Sciences, National Research Council (CNR-ISMAR), Venice, Italy.

5. UCD School of Mathematical Sciences, University College Dublin, Belfield, Dublin 4, Ireland.

1  **Abstract**

2   Observations show that maximally-steep breaking water wave crest speeds are much slower than
3   expected. We report a wave-crest slowdown mechanism generic to *unsteady* propagating deep water
4   wave groups. Our fully nonlinear computations show that just prior to reaching its maximum height, each
5   wave crest slows down significantly and either breaks at this reduced speed, or accelerates forward
6   unbroken. This finding is validated in our extensive laboratory and field observations. This behavior
7   appears to be generic to unsteady dispersive wave groups in other natural systems.

9  **Introduction**

10   Nonlinear wave groups occur in a wide range of natural systems, exhibiting complex behaviors
11   especially in focal zones where there is rapid wave energy concentration and possible 'wave-breaking'.
12   The incompletely-understood interplay between dispersion, directionality and nonlinearity presents a
13   significant knowledge gap presently beyond analytical treatment. Here, we investigate maximally-steep,

14    deep-water wave group behaviour, but the findings appear relevant to dispersive nonlinear wave motion

15    in many other natural systems.

16    In the open ocean, wind forcing generates waves that can steepen and break conspicuously as

17    whitecaps, strongly affecting fundamental air-sea exchanges, including greenhouse gases. This has

18    stimulated recent interest in measuring whitecap properties spectrally. While accurately measuring

19    wavelengths of individual breakers is difficult, measuring whitecap *speeds* can provide a less direct but

20    more convenient method: since a whitecap remains attached to the underlying wave crest during active

21    breaking. The dispersion relation from Stokes' classical deep water wave theory discussed below (Stokes

22    [1]) conventionally provides the wavelength from the observed whitecap speed (Phillips [2]).

23    Stokes' theory was developed for a *steady*, uniform train of two-dimensional (2D) non-linear, deep-

24    water waves of small-to-intermediate mean steepness $ak$ (=2π×amplitude/wavelength), for which the

25    intrinsic wave speed $c$ increases slowly with $ak$:

26    $$c = c_0[1+1/2\ (ak)^2+\text{higher order terms in } (ak)]^{1/2} \tag{1}$$

27    where $c_0$ is the wave speed for linear (infinitesimally-steep) waves. Extending (1) computationally to

28    maximally-steep, steady waves (Longuet-Higgins [3]), $c$ approaches $1.1c_0$. Thus, increased wave

29    steepness has long been associated with *higher* wave speeds.

30    Natural wind-waves comprise a spectrum of modes interacting on different scales, producing evolving

31    wave-group patterns rather than steady, uniform wavetrains (Longuet-Higgins [4]). Here we investigate

32    the 'dominant' waves, i.e. those with the largest spectral amplitudes after filtration of higher-wavenumber

33    components. Within a group, each advancing dominant wave gradually changes its height and shape,

34    characterized by slow forward and backward leaning of the crests (Tayfun [5]), also transiently becoming

35    the tallest wave. This tallest wave may break, or else decrease in height while advancing unbroken

36    towards the front of the group.

37    In this context, previous deep-water breaking wave laboratory studies (Rapp and Melville [6]; Stansell

38    and McFarlane [7]; Jessup and Phadnis [8]) suggest that breaking-crest speeds are typically O(20%)

lower than expected from linear-wave theory, contrary to the expectation from (1) that steeper breaking waves should propagate faster. Understanding this paradoxical crest slowdown behaviour is central to both refining present knowledge on water-wave propagation and dynamics, and optimal implementation of Phillips' spectral framework for breaking waves (Phillips [2]; Kleiss and Melville [9]; Gemmrich et al. [10]).

Historically, an appreciable literature has developed on non-breaking, focusing, deep-water, nonlinear wave packets. However, only the studies of Johannessen and Swan ([11], [12]) identified crest slowdown at focus, reporting an O(10%) crest-speed slowdown relative to its linear-theory prediction. To understand the underlying physics, the present study investigates how very steep unsteady, non-periodic, deep-water wave groups propagate when frequently-assumed theoretical constraints are relaxed, including steady-state, spatially-uniform or slowly-varying, weakly-nonlinear wavetrain behavior. Our goal was to investigate initial breaker speeds, hence it was crucial to track changes, up to the point of breaking initiation, in dominant wave-crest speeds within evolving nonlinear wave groups.

**Methodology and results**

No presently-available analytic theory can predict the evolution of fully-nonlinear, deep-water wave groups. Our primary research strategy utilized simulations from a fully-nonlinear, 3D numerical wave code, validated against results from our innovative laboratory and ocean-wave observations.

Our simulations were generated using a numerical wave tank (Grilli et al. [13]). This boundary element code simulates fully-nonlinear potential flow theory and is able to model extreme water waves to the point of overturning. A programmable wave paddle produces a specific 2D or 3D chirped wave-group structure comprising a prescribed number of carrier waves with given initial amplitudes, wavenumbers, frequencies and phases. This shapes the spatial and temporal bandwidths characterizing the group structure and its spectrum. For the simulations, including the 2D example below, the paddle followed the displacement-motion equation (3) described in Song and Banner [14], with N=5, 7 and 9. We also investigated corresponding laterally-converging 3D chirped packet cases with 10- and 25-wavelength

65   focal distances. In this study, breaking occurred predominantly as sequential spilling events with

66   occasional local plunging. The complementary wave-basin experiments described below also included

67   comparable bimodal, modulating nonlinear wave packets specified by equation (2) in [14]. The half-

68   power bandwidths were O(8) times broader than investigated in [9].

69   Figure 1(a) shows the complex growth behavior experienced by all dominant wave crests evolving

70   within a representative 2D nonlinear, non-breaking wave group. The initial steepest wave decays and is

71   replaced by the following growing wave, which grows modestly, then slows down and is replaced by the

72   annotated faster-growing crest, which evolves to its maximum height and decays. As each new crest

73   develops, it grows (A-B-C) then slows down and attenuates (C-D), then accelerates (D-E) back to its

74   original speed while advancing towards the front of the group.

75   Figure 1(b) shows the spatial wave profile in greater detail at the evolution times A-E in Figure 1(a).

76   The dominant wave grows asymmetrically, initially leaning forward as it steepens within the group. In the

77   absence of breaking, the steepest wave advances leaning forward, relaxing back to symmetry near its

78   maximum height (the focal point), then leans backwards past the maximum elevation. Forward-leaning

79   crests are accompanied by backward leaning troughs, and vice-versa. This leaning is a *generic* feature of

80   each crest in natural, unsteadily-evolving dispersive nonlinear water wave groups (Tayfun [5]).

81   Relative to the speed of a classical (symmetrical) Stokes wave, significant crest (and trough) speed

82   changes accompany the leaning, measured by tracking the (horizontal) speed of a given wave-crest profile

83   in space and time. The generic crest-speed slowdown is identified in Figure 1(c) by the steeper slope of

84   the displacement-time curve between B and D relative to the indicated *linear* wave trajectory (speed $c_0$),

85   where $c_0$ is the speed of the spectral peak determined from the computed wave-packet dispersion relation.

86   The actual speed reduction relative to $c_0$ is 18%. This lasts about one wave period, with a spatial extent of

87   about one wavelength. Figure 1(d) shows a typical trajectory when crest speed is plotted against local

88   crest steepness $s_c = a_c k_c$, where $a_c$ is the time-dependent crest height above mean-water level and the

89   corresponding local wavenumber $k_c$ is defined as $\pi$ divided by the local zero-crossing separation spanning

90    the given crest. Introducing $s_c$ was necessary to describe the complexity of *unsteady* nonlinear wave crest

91    behaviour, and was easily computed for Stokes waves for the crest-speed comparison shown.

92      The significant departure of the crest speed versus crest steepness trajectory for waves in unsteady

93    wave groups compared with the classical Stokes for steady wavetrain prediction underpins the central

94    findings in this study. In this example, the maximum crest steepness marginally precedes the slowest

95    crest speed, with the trajectory looping counter-clockwise about this point. This trajectory is not generic,

96    since other simulated cases and the experimental curve of Figure 2(b) below showed clockwise looping.

97    Further studies are needed to explain this effect. Also, as seen in Figure 1(d), the asymmetry of the

98    dominant wave shape near its maximum steepness results in different crest speeds for growing and

99    decaying crests of the same steepness. Note that the local peak in crest speed between A and B at $s_c \approx 0.2$ is

100   an artefact of our crest-tracking algorithm, resulting from the complex crest transition seen in Figure 1(a)

101   at ($x/L \sim 4$, $t/T \sim 13.5$) when the detected crest location jumps abruptly from the receding crest to the newly-

102   developing crest.

103      The above discussion was for 2D waves, but laterally-focused (3D) wave fronts in both our simulations

104   and wave-basin investigation (described below) show similar leaning and crest-slowdown behavior, with

105   the subsequent breaker-crest speed initiated at $\sim 0.8 c_0$.

106      Relative to classical ocean-wave speeds, our model results for the speeds of the left-hand and right-

107   hand zero-crossings spanning the tallest wave shows that their average remains close to the linear wave

108   speed $c_0$ (Fig. 1(c)), with modest local fluctuations of (+7% to -1%). Hence, aside from the strong

109   unsteady leaning crest and trough motions, the waves propagate largely as expected from Stokes theory.

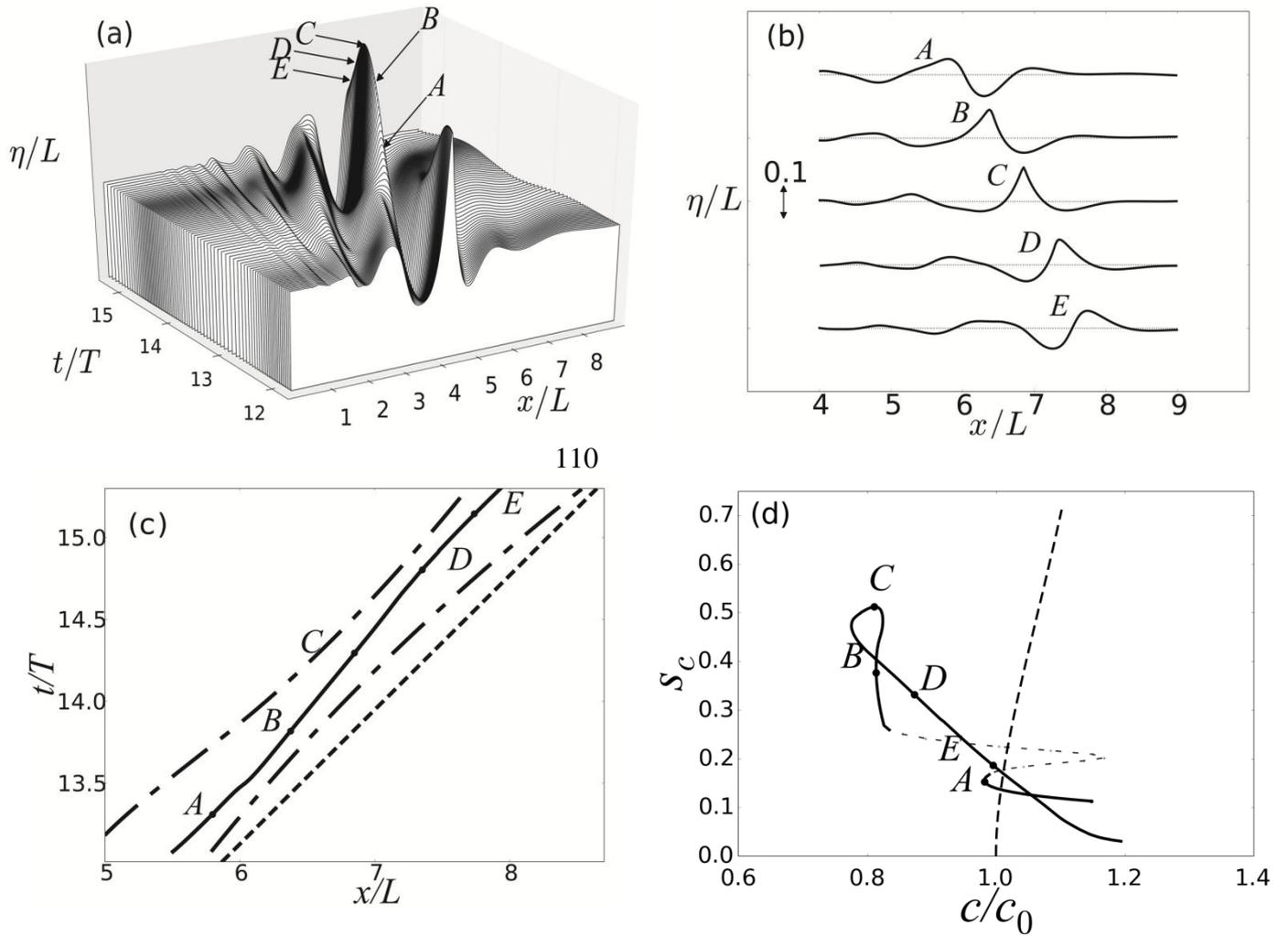

Figure 1. (a) space-time evolution diagram of a non-breaking 2D chirped wave group, moving toward the right, showing the decay of the initial tallest crest, growth of the following tallest crest and complex transitions of other developing crests. Wave properties at annotated times A-E are shown in panels (b), (c) and (d). $T$ and $L$ are reference carrier-wave period and wavelength scales. (b) tallest crest shapes at evolution times A-E, showing crest transition from forward-leaning through symmetry to backward-leaning (c) horizontal location of the tallest crest (solid line) versus time. The steeper slope between B and D shows the crest-speed reduction relative to $c_0$ (dotted line). Horizontal locations versus time of the two adjacent zero-crossings (long-short dashed lines) are also shown. (d) trajectory of the corresponding tallest crest speed $c$, normalized by $c_0$, against local crest steepness $s_c$ defined in the text. Stokes theory prediction (1) is shown in terms of $s_c$ (dashed line) for comparison. The apparent crest-speed surge at $s_c \approx 0.2$ is spurious, as explained in the text.

123

*Breaking onset and speed*

124

125    If the tallest wave in the group proceeds to break rather than recur, our simulations found that breaking

126    onset occurs when this wave attains maximum steepness and close to its minimum crest speed. This can

127    certainly explain why initial breaking wave crest speeds are observed to be O(80%) of the linear carrier-

128    wave speed (Rapp and Melville [6]; Stansell and McFarlane [7]; Jessup and Phadnis [8]). This behavior

129    was found in all our simulations and verified in our laboratory measurements (see Figure 2(b)).

130

*Is the crest slowdown a nonlinear effect?*

131

132    Insight on this key question is available from previous linear and weakly-nonlinear theory. For

133    uniform, deep-water, linear gravity wavetrains, the carrier wave speed $c_0$ follows from the dispersion

134    relation $\omega = (gk)^{1/2}$ and $c_0 = \omega_0/k$. However, narrow-band wave groups are characterized by non-

135    uniformity in both space and time. Correct to $O(\nu^2)$, a local frequency can be defined (Chu and Mei [15])

136    as

137
$$\omega = (gk)^{1/2} - \beta\, a_{xx}/(ak) \qquad (2)$$

138    where $\nu$ is a characteristic spectral bandwidth, $\beta = dc_g/dk$, $c_g = d\omega_0/dk$ is the linear group velocity and $a(x,t)$

139    is the wavetrain envelope that satisfies the linear Schrödinger equation (Mei [16]). The associated local

140    phase speed is given approximately by

141
$$c \approx c_0 - \beta\, a_{xx}/(ak^2). \qquad (3)$$

142    Equations (2), (3) and the associated relationships above show that $c$ varies along the group and in time.

143    Since $\beta < 0$, $c$ attains its lowest value at the envelope maximum, where the largest crest occurs ($a_{xx} < 0$).

144    The other crests and troughs in the group also experience similar local speed variations.

145     Furthermore, for dispersive, weakly-nonlinear unsteady wave groups, we find that the slowdown effect

146     due to dispersion is counterbalanced by the increase in phase speed due to nonlinearity [Equation (1)],

147     limiting the phase-velocity slowdown within the group (Fedele [17]).

148     Our focus on breaking-crest slowdown for large wave steepness approaching breaking onset is beyond

149     conventional analysis methodologies. We now validate our fully-nonlinear numerical simulation findings

150     on wave-crest slowdown against laboratory and open-ocean measurements.

151

152     *Wave basin measurements*

153     Complementary experiments were performed in a 27m x 7.75m wave basin with 0.55m water depth.

154     Wave groups were generated at one end of the basin by a computer-controlled wave-generator comprising

155     13 bottom-cantilevered, flexible-plate segments. Lateral focusing was achieved by suitably setting the

156     phase of each segment (Dalrymple [18]). A 95% absorbing beach minimized end reflections. Heights of

157     evolving wave groups matching the simulations were measured to within ±0.5mm by a traversable in-line

158     array of nine wave-wire probes spanning one wavelength. $c_0$ was calculated using linear theory from the

159     spectrally-weighted wave frequency of the wave probe closest to the wave-generator.

160     Identified wave crests were tracked between the wave probe signals, their motion interpolated using

161     cubic splining and their crest speeds extracted. An overhead one-megapixel videocamera imaged the

162     breaking crests at 100 Hz. The imagery, corrected for lens and mounting distortion, was transformed onto

163     a regular grid, sequential leading-edge location and lateral extent data were extracted for each breaker and

164     their speeds determined.

165     Figure 2(a) shows surface profiles measured at evolution times A-E, with breaking initiation near C, for

166     a modulating 5-wave, bi-modal breaking case. Figure 2(b) shows its crest speed trajectory. Also shown is

167     the ensemble-mean trajectory for spilling-breaker speeds in the measured ensemble of 240 modulational

168     and chirped 2D and 3D cases. The measurement resolution enabled resolving the crest leaning and

169     slowing at the maximum surface elevation (C). Crest-speed oscillations observed for smaller-steepness

170     waves (e.g. at B), are the same crest-leanings, but occurring earlier as the crest moves through the wave group. This figure confirms the reduced speeds of crests preceding breaking onset and the accompanying generic breaker slowdown.

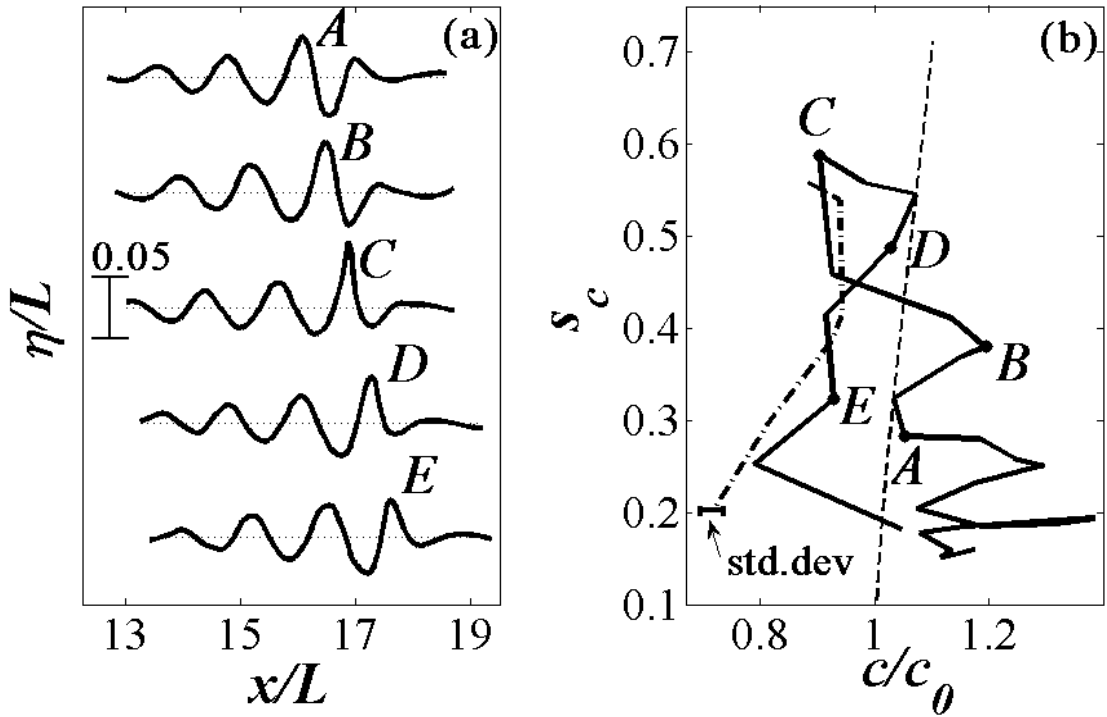

Figure 2. (a) measured surface profiles of a 5-wave, bi-modal wave packet, at times A-E, with breaking initiation at C. (b) corresponding trajectory of normalized crest speed $c/c_0$ of the tallest wave, against local crest steepness $s_c$. The ensemble-mean breaker speed trajectory is shown by the dot-dash line. The dashed line shows Stokes' prediction. Reference wave scales were $L$=1.09m, $T$=0.836 sec.

*Open ocean observations*

Our *Wave Acquisition Stereo System (WASS)* was deployed at the *Acqua Alta* oceanographic tower 16 km offshore from Venice in 17m water depth (Fedele et al. [19]; Benetazzo et al. [20]). *WASS* cameras were 2.5m apart, 12.5m above sea level at 70° depression angle, providing a trapezoidal field-of-view with sides increasing from 30m to 100m over a 100m fetch. The mean windspeed was 9.6 ms[-1] with a 110 km fetch. The uni-modal wave spectrum had a significant wave height $H_s$=1.09m and dominant period $T_p$=4.59s. Most observed crests were very steep, with sporadic spilling breaking. We describe results using 21,000 frames captured at 10 Hz.

The speeds $c$ of crests reaching maximum local steepness within the imaged area were estimated using a crest-tracking methodology, as in the wave-basin measurements. The data were filtered above 1.5 Hz to remove short riding waves. Sub-pixeling reduced quantization errors in estimating the local 3D crest position from the surface-displacement time series spaced along the wave-propagation direction. The local reference $c_0$ was calculated from the peak frequency of the short-term Fourier spectrum of a time series of duration $D$ centered at the crest event, using $D$=120sec as a suitable record length and Doppler-corrected for the in-line 0.20 ms$^{-1}$ mean current. We analyzed 200 dominant local wave crests with elevations $\eta$>0.3$H_s$ and local crest steepness $s_c$>0.3$(s_c)_{max}$ using the observed $(s_c)_{max}$=0.45, and determined ~12,000 evolving crest speeds from a 60-point spatial grid, with 0.5m spacing along the wave-propagation direction.

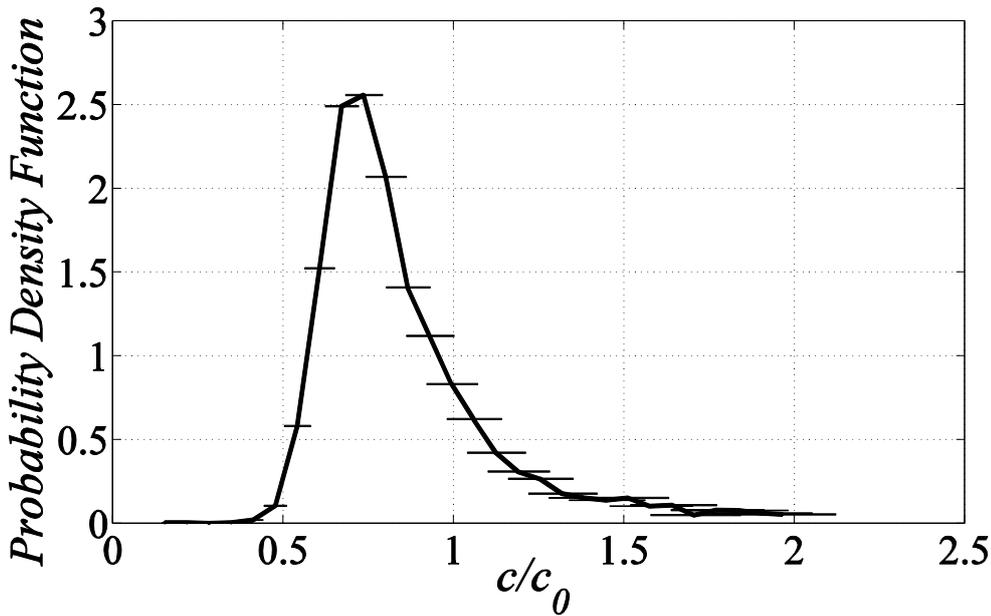

Figure 3. Probability density function of normalized crest speed $c/c_0$ for all crests transitioning through a maximum local crest steepness, from a 35-minute *WASS* stereo-video sequence from an ocean tower. Note the tall peak at $c/c_0 \sim 0.75$. Local standard error bounds are indicated.

218    Values of $D$ and $\eta$ were chosen so that the empirical probability density function (pdf) of $c/c_0$ was

219    insensitive to changes in these parameters. Figure 3 shows the pdf, which peaks at close to $0.75c_0$. Values

220    for $c/c_0 > 1.5$ (7% of the total ensemble) are outliers with >15% uncertainty in estimating $c_0$ and crest

221    location. This figure highlights the observed crest slowdown, consistent with the nonlinear simulations

222    and experiments described above.

223

224    **Discussion and conclusions**

225    Our study provides fundamental new insights into the behavior of chirped, bi-modal and open-ocean

226    unsteady steep, deep-water nonlinear wave groups. We found that as carrier waves reach maximum

227    steepness, their crests decelerate strongly (O(20%)), which results from unsteady crest sloshing modes

228    arising from the complex interplay between nonlinearity and dispersion. This behaviour departs markedly

229    from the speed increase with wave steepness predicted by steady-wavetrain theory.

230    Our findings have significant, broader consequences. For ocean waves, they explain the puzzling

231    (O(20%)) reduced initial speed of breaking-wave crests, central to assimilating whitecap data accurately

232    into sea-state forecast models. Parameterizations of air-sea fluxes of momentum and energy, which

233    depend on the square and cube of the sea-surface velocity, may be modified appreciably. Atmospheric

234    and oceanic internal waves, (Helfrich and Melville [21]), should also experience similar effects to those

235    described here. As noted above, even weakly-nonlinear, unsteady dispersive water-wave groups described

236    by the nonlinear Schrödinger equation (NLSE) (Zakharov [22]) exhibit crest slowdown. The NLSE is

237    commonly used to describe wave phenomena in other natural systems (e.g. geophysical flows (Osborne

238    [23]), nonlinear optics (Kibler et al. [24], amongst others). Exploring implications of the present findings

239    should provide refined insights when the wave-group nonlinearity and bandwidth are beyond the validity

240    of the NLSE.

241

242

273

## Acknowledgements

275    Financial support is gratefully acknowledged for XB, MA, MB and WP from the Australian Research

276    Council through their support of Discovery Projects DP0985602, DP120101701. Financial support for

277    MB is also gratefully acknowledged from the National Ocean Partnership Program, through the U.S.

278    Office of Naval Research (Grant N00014-10-1-0390). The WASS experiment at *Acqua Alta* was

279    supported by Chevron (CASE-EJIP Joint Industry Project #4545093). FD was partially supported by ERC

280    under the research project ERC-2011-AdG 290562-MULTIWAVE and SFI under the programme ERC

281    Starter Grant - Top Up, Grant 12/ERC/E2227.

282    In this paper, XB identified and quantified the crest slowdown effect in his steep nonlinear wave

283    computations. MB made the central association with breaker slowdown and was the architect of this

284    paper, coordinating the scientific effort.  FF identified that the crest slowdown also occurs in linear wave

285    groups. He revealed that wave group bandwidth and unsteadiness were crucial aspects of the crest

286    slowdown, using linear/nonlinear narrow-band wave theory.  MA performed the suite of laboratory wave

287    basin experiments relating crest velocity and breaker speed. AB deployed the WASS, processed stereo

288    data and contributed jointly with FF the WASS ocean wave crest speed analysis. WP and FD made

289    ongoing incisive intellectual contributions.